\documentstyle[11pt,newpasp,twoside,epsf]{article}
\def\edcomment#1{\iffalse\marginpar{\raggedright\sl#1\/}\else\relax\fi}
\reversemarginpar

\begin{document}
\title{Gravity, Turbulence, and Star Formation}
\author{Bruce Elmegreen}
\affil{IBM Research Division, T.J. Watson Research
    Center, PO Box 218, Yorktown Hts., NY, 10598, USA,
    bge@watson.ibm.com}

\begin{abstract}
The azimuthal power spectra of optical emission from star formation
and dust in spiral galaxies resembles the azimuthal power spectra of HI
emission from the LMC.  These and other power spectra of whole galaxies
all resemble that of velocity in incompressible Kolmogorov turbulence.
The reasons for this are unknown but it could be simply that star
and cloud formation are the result of a mixture of processes and each
gives a power spectrum similar to Kolmogorov turbulence, within the
observable errors.  The important point is that star and cloud formation
are not random but are correlated over large distances by forces that
span several orders of magnitude in scale. These forces are probably
the usual combination of self-gravity, turbulence, and compression from
stellar winds and supernovae, but they have to work in concert to create the
structures we see in galaxies.  In addition, the
identification of flocculant spirals with swing amplified
instabilities opens the possibility that
a high fraction of turbulence in the ISM is the result
of self-gravity. \end{abstract}

\noindent In Star Formation in the Interstellar Medium: A 60th birthday celebration
for David Hollenbach, Chris McKee, and Frank Shu, Ed. F. Adams, D. Johnstone, D. Lin and E. Ostriker,
Astronomical Society of the Pacific (PASP Conference series), in press

\section{Introduction}
Many recent studies of ISM gas have shown correlated behavior over
a wide range of scales. These observations include power spectra
of HI emission in our Galaxy (Dickey et al. 2001), the SMC
(Stanimirovic et al. 1999) and the LMC (Elmegreen, Kim, \&
Staveley-Smith 2001), in addition to many other gas tracers,
including molecules (St\"uzke et al. 1998) and dust (Stanimirovic
et al. 2000).  A review of these and other correlated properties
of the ISM is in Elmegreen \& Scalo (2004).

We are interested in correlations specifically related to star
formation. Algorithms using nearest-neighbors and cell counting
methods find star cluster hierarchies up to kpc scales in many
galaxies (Feitzinger \& Braunsfurth 1984; Feitzinger \& Galinski
1987; Battinelli, Efremov, \& Magnier 1996; Elmegreen \& Salzer
1999; Heydari-Malayeri et al. 2001; Pietrzynski et al. 2001).
Unsharp masks of optical images also show self-similar structure
from tens of pc to multi-kpc scales (Elmegreen \& Elmegreen 2001),
as do auto-correlation functions (Harris \& Zaritsky 1999; Zhang,
Fall, \& Whitmore 2001), which go up to about a kpc. Self-similar
structure also goes down to sub-parsec scales inside nearby
regions (Testi et al. 2000), making the definition of a unique
``cluster'' sometimes difficult.

These patterns are apparently present as long as the regions are
younger than a crossing time, regardless of scale (Elmegreen
2000). After this, random and orbital motions smooth out the
stellar positions.

\section{New Results}

We have recently found that flocculent spiral arms in galaxies
have a power law power spectrum too (Elmegreen et al. 2003a,b).
This is observed in azimuthal scans of optical intensity in
several passbands, and in scans along density wave spiral arms.
The azimuthal scans also show the spiral density waves, which
are the most obvious features seen directly by the eye, but the
density waves are
confined to the lowest few wavenumbers in power spectra, as predicted by
density wave theory (Bertin et al. 1989).  At higher wavenumbers, grand
design spiral galaxies look the same as flocculent galaxies in
their power spectra.

\begin{figure}
\plotone{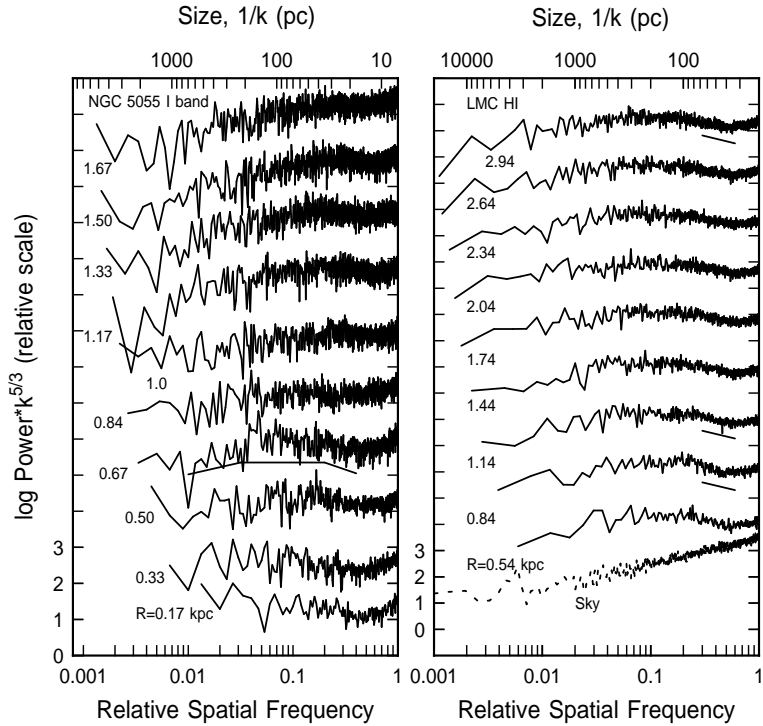} \caption{A comparison between
the power spectra of optical emission from the flocculent galaxy
NGC 5055 and HI emission from the LMC. Power spectra have been
multiplied by $k^{5/3}$ to flatten any portions that have
the same slope as Kolmogorov turbulence.}
\end{figure}

Figure 1 shows a comparison between the power spectra of the azimuthal
scans of the optical emission from the flocculent galaxy NGC 5055,
using I band data from HST, and the HI emission from the LMC.  To make
a slope of $-5/3$ easier to see, we multiplied each power spectrum by
$k^{5/3}$ so that a Kolmogorov spectrum of $k^{-5/3}$ would look flat.
Each scan corresponds to a different radius, as indicated, and a scan
through the sky around the LMC, where the HI emission is very weak, is
shown at the bottom right.  The power spectra of these two galaxies look
remarkably similar, and the power spectra at each radius look similar
too, except for the increasing influence of noise with radius, which
makes the spectra look like the sky spectrum.  There is a relatively
flat part in the center of the power spectra at mid-wavenumber range,
where the power spectra have the Kolmogorov slope. Also for both galaxies,
the power spectra systematically dip down and up at high $k$.  This dip
could be the result of unresolved structure: stars and unresolved clusters
in the case of NGC 5055 and perhaps unresolved HI clouds in the case of
the LMC. It is also possible that the dip in some cases is the result
of a transition from 2D structure on large scales (on the left) to 3D
structure on scales smaller than the galaxy thickness at the highest $k$.
The line segments in NGC 5055 show slopes of $2/3$, $0$, and $-1$, and
the line segments in the LMC also have a slope of $-1$.  The transition
from 2D to 3D is expected to change the slope in the power spectrum by 1,
as indicated by the change from $0$ to $-1$ in this figure.  Models of
the HI power spectra are in Elmegreen et al. (2001) and models of the
NGC 5055 power spectra are in Elmegreen et al. (2003b).

The power spectra found in optical images have close to a
Kolmogorov slope, although the uncertainties are large because
foreground stars introduce distortions that have to be modelled
(Elmegreen 2003b). Nevertheless, this result suggests that the
standard model for the origin of flocculent arms, namely swing
amplified gravitational instabilities (e.g., Toomre \& Kalnajs
1991; Wada, Meurer, \& Norman 2002), makes structure not only on
the characteristic scale of the instability, which is $k^{-1}\sim
a^2/\left(\pi G\Sigma\right)\sim$ kpc for wavenumber $k$, velocity
dispersion $a$ and mass column density $\Sigma$, but also on much
smaller scales, down to the limit of optical resolution (which is
essentially the limit where the power spectra are dominated by
individual stars, unresolved clusters, etc.).

If gravitational instabilities are directly or indirectly related
to most of this structure, and perhaps even to the resulting star
formation that is observed in the optical surveys, then these
instabilities could be an important energy source for ISM motions.
Widespread instabilities have an available power density that is
approximately the energy density of the ISM multiplied by the
instability growth rate of $\sim\pi G\Sigma/a\sim1/30$ My for
small Toomre $Q$.  This power density is $\sim10^{-27}$ erg
cm$^{-3}$ s$^{-1}$ if the conversion to motion is highly
efficient. It is an order of magnitude lower than the power
density available from supernovae, including their low (0.1)
efficiency (Mac Low 2002), but may dominate the energy sources on
the largest scales (Crosthwaite, Turner, \& Ho 2000).

Numerical simulations of galaxy disks do indeed find reasonable
rms velocities from repetitive gravitational instabilities
(Thomasson, Donner, \& Elmegreen 1991; Fuchs \& von Linden 1998;
Bertin \& Lodato 2001; Wada \& Norman 2001; Wada, Meurer, \&
Norman 2002; Huber \& Pfenniger 2001).  The resulting structures
may be scale-free as well (Huber \& Pfenniger 2002).

\section{Implications for turbulence}

Sources of turbulent energy in the ISM span a wide range of
scales. In a compilation by Norman \& Ferrara (1996), the largest
scale was powered by superbubbles, but it may be that
gravitational instabilities are important on large scales too.
Studies of the largest HI clouds in the inner galaxy suggested
long ago (Elmegreen \& Elmegreen 1987) that virialized motions
account for $\sim0.5$ of the random kinetic energy in midplane HI.
The largest CO clouds are virialized too (Heyer et al. 2001), and
if they also represent $\sim0.5$ of the total molecular mass,
according to the cloud mass spectrum, then half of that internal
kinetic energy is gravitational also.  Because most molecular
clouds are parts of giant HI clouds, at least in the Carina arm
(Grabelsky et al. 1987), even a good fraction of the GMC
cloud-to-cloud motions could be virialized motions.

The shocks of spiral density waves could stimulate turbulence also.
Zhang et al. (2001) suggest that a spiral wave has driven
turbulence in the Carina molecular clouds because the
linewidth-size relation is not correlated with distance from the
stellar energy sources. The mechanism for this
driving is unknown but it could be gravitational if the spiral
shock triggers an instability (Kim \& Ostriker 2001).

The total contribution from self-gravity to the turbulent motion
of the ISM is not known, but it could be second in importance next
to supernovae. A detailed assessment requires a re-evaluation of
both sources considering their distributions, the stratification of the
disk, and the efficiency of conversion of directed motions into
turbulence.

\section{Conclusions}

Power spectra of galactic-scale structures have power law forms
similar to that of velocity in incompressible Kolmogorov turbulence.
The observations permit a range of $\pm0.2$ around the Kolmogorov
slope, and this large range, along with the lack of theoretical
predictions about what the density structure should be in compressible,
self-gravitating, sheared MHD turbulence, prevent the use of these power
spectra as diagnostics for definite physical processes. For example, ISM
structures are a composite of shells, comets, self-gravitating clumps,
and turbulence-compressed regions, and all of these have about the same
power spectra.

What is important is that there is a power law power spectrum at all.
This occurs for a wide range of structures ranging from nested shells
in the LMC, to dust spirals in galactic nuclei, to flocculent star
formation arms in optical galaxies. These structures are self-similar
over several orders of magnitude in scale.  This observation argues
for coherent formation processes, although these processes may involve
various mixtures of forces in different regions. The observations
also suggest that all of these forces contribute to ISM tubulence.


\begin{thebibliography}{}


\bibitem[]{} Battinelli, P., Efremov, Y., \& Magnier, E.A. 1996, A\&A, 314, 51

\bibitem[]{} Bertin, G., Lin, C. C., Lowe, S. A., \& Thurstans, R. P. 1989,
ApJ, 338, 78

\bibitem[]{} Bertin, G. \& Lodato, G. 2001, A\&A, 370, 342

\bibitem[]{} Crosthwaite, L.P., Turner, J.L., \& Ho, P.T.P. 2000, AJ, 119, 1720

\bibitem[]{} Dickey, J.M., McClure-Griffiths, N.M., Stanimirovic, S.,
Gaensler, B.M, \& Green, A.J, 2001, ApJ, 561, 264

\bibitem[]{} Elmegreen, B.G. 2000, ApJ, 530, 277

\bibitem[]{} Elmegreen, B.G., \& Elmegreen, D.M. 1987, ApJ, 320, 182

\bibitem[]{} Elmegreen, B.G., Kim, S., Staveley-Smith, L. 2001, ApJ, 548,749

\bibitem[]{} Elmegreen, B.G., \& Elmegreen, D.M. 2001, AJ, 121, 1507

\bibitem[]{} Elmegreen, B.G. \& Scalo, J.M. 2004, ARAA, in press.

\bibitem[]{} Elmegreen, B.G., Elmegreen, D.M., Leitner, S.N.  2003a, ApJ, 590, 271

\bibitem[]{} Elmegreen, B.G., Leitner, S.N., Elmegreen, D.M., Cuillandre, J.-C.
2003b, ApJ, 593, 333

\bibitem[]{} Elmegreen, D.M., \& Salzer, J.J. 1999, AJ, 117, 764

\bibitem[]{} Feitzinger, J.V., \& Braunsfurth, E. 1984, A\&A, 139, 104

\bibitem[]{} Feitzinger, J. V., \& Galinski, T. 1987, A\&A, 179, 249

\bibitem[]{} Fuchs, B., \& von Linden, S. 1998, MNRAS, 294, 513

\bibitem[]{} Grabelsky, D.A., Cohen, R.S., Bronfman, L.,
Thaddeus, P., May, J. 1987, ApJ, 315, 122

\bibitem[]{} Harris, J., \& Zaritsky, D. 1999, AJ, 117, 2831

\bibitem[]{} Heydari-Malayeri, M., Charmandaris, V., Deharveng, L., Rosa, M.R.,
Schaerer, D., \& Zinnecker, H. 2001, A\&A, 372, 495

\bibitem[]{} Heyer, M. H., Carpenter J. M., Snell, R. L. 2001, ApJ, 551, 852

\bibitem[]{} Huber, D. \& Pfenniger, D. 2001, A\&A, 374, 465

\bibitem[]{} Huber, D., \& Pfenniger, D. 2002, A\&A, 386, 359

\bibitem[]{} Kim, W.-T., \& Ostriker, E.C. 2001, ApJ, 559, 70

\bibitem[]{} Mac Low, M. 2002, astroph/0211616

\bibitem[]{} Norman, C.A., Ferrara, A. 1996, ApJ, 467, 280

\bibitem[]{} Pietrzynski, G. Gieren, W., Fouqu\'e, P. \& Pont, F.
2001, A\&A, 371, 497

\bibitem[]{} Stanimirovic, S., Staveley-Smith, L., Dickey, J.M.,
Sault, R.J., \& Snowden, S.L. 1999, MNRAS, 302, 417

\bibitem[]{} Stanimirovic, S., Staveley-Smith, L., van der Hulst, J.M.,
Bontekoe, TJ.R., Kester, D.J.M., Jones, P.A. 2000, MNRAS, 315, 791

\bibitem[]{} St\"utzki, J., Bensch, F., Heithausen, A., Ossenkopf, V.,
Zielinsky, M. 1998, A\&A, 336, 697

\bibitem[]{} Testi, L., Sargent, A.I., Olmi, L., \& Onello, J.S., 2000, ApJ, 540, L53

\bibitem[]{} Thomasson, M., Donner, K.J, \& Elmegreen, B.G. 1991,
A\&A, 250, 316

\bibitem[]{} Toomre,~A., \& Kalnajs,~A.~J. 1991, in Dynamics of Disk
Galaxies, ed. B. Sundelius, University of Chalmers, p. 341

\bibitem[]{} Wada,~K., \& Norman,~C.~A. 2001, ApJ, 547, 172

\bibitem[]{} Wada, K., Meurer, G., \& Norman, C.A. 2002, ApJ, 577,
197

\bibitem[]{} Zhang, Q., Fall, S.~M., \& Whitmore, B.~C. 2001, ApJ, 561, 727

\bibitem[]{} Zhang, X., Lee, Y., Bolatto, A., Stark, A.A. 2001,
ApJ, 553, 274


\end{thebibliography}
\end{document}